\documentclass[16pt,twocolumn,amsmath,amssymb]{revtex4}
\usepackage{graphicx}
\usepackage{dcolumn}
\usepackage{bm}
\usepackage{amsmath}
\usepackage{amssymb}
\usepackage{amsmath,amsthm,amssymb,amscd}
\usepackage{latexsym}
\usepackage{indentfirst}
\usepackage{subfigure}
\usepackage{graphicx}

\date{\today}

\begin{document}

\title{
\bf Growth and shortening of microtubules: a two-state model approach
}

\author{Yunxin Zhang}\email[Email: ]{xyz@fudan.edu.cn}
\affiliation{
Laboratory of Mathematics for Nonlinear Science, Centre for Computational System Biology,
School of Mathematical Sciences, Fudan University, Shanghai 200433, China.
}

\begin{abstract}
In this study, a two-state mechanochemical model is presented to describe the dynamic instability of microtubules (MTs) in cells. The MTs switches between two states, assembly state and disassembly state. In assembly state, the growth of MTs includes two processes: free GTP-tubulin binding to the tip of protofilament (PF) and conformation change of PF, during which the first tubulin unit which curls outwards is rearranged into MT surface using the energy released from the hydrolysis of GTP in the penultimate tubulin unit. In disassembly state, the shortening of MTs includes also two processes, the release of GDP-tibulin from the tip of PF and one new tubulin unit curls out of the MT surface. Switches between these two states, which are usually called rescue and catastrophe, happen stochastically with external force dependent rates. Using this two-state model with parameters obtained by fitting the recent experimental data, detailed properties of MT growth are obtained, we find that MT is mainly in assembly state, its mean growth velocity increases with external force and GTP-tubulin concentration, MT will shorten in average without external force. To know more about the external force and GTP-tubulin concentration dependent properties of MT growth, and for the sake of the future experimental verification of this two-state model, eleven {\it critical forces} are defined and numerically discussed.

\end{abstract}

\maketitle

\section{Introduction}
In eukaryotic cells, microtubules (MTs) serve as tracks for motor proteins \cite{Frank1995, Schnitzer1997, Vale2003, Schliwa2003, Sperry2007, Kolomeisky2007}, give shape to cells, and form rigid cores of organelles \cite{Cooper2000, Howard2001, Howard2006, Howard2009}. They also play essential roles in the chromosome segregation \cite{Westermann2005, Miranda2005, Grishchuk2005, Westermann2006, Franck2007, McIntosh2008, Powers2009, Gao2010}. During cell division, MTs in spindle constantly grow and shorten by addition and loss of enzyme tubulin (GTPase) from their tips, then the attached duplicated chromosomes are stretched apart (through two kinetochores) from one another by the opposing forces (produced by MTs based on different spindles). Recently, many theoretical models have been designed to understand the roles of MTs during chromosome segregation \cite{Hill1985, Molodtsov2005, Salmon2005, Grill2005, Efremov2007, Armond2010, Asbury2011}. One essential point to understand how MTs help chromosome segregation during cell division is to know the mechanism of MT growth and shortening. In this study, inspired by the mechanochemical model for molecular motors \cite{Fisher2001}, the GTP-cap model and catch bonds model for MT \cite{Howard2001, Akiyoshi2010}, a two-state mechanochemical model will be presented.

Electron microscopy indicates MT is composed of $n$ parallel protofilaments (PFs, usually $12\le n\le 15$ and $n=13$ is used in this study) which form a hollow cylinder \cite{Cooper2000, Howard2001, Bray2001}. Each PF is a filament that made of head-to-tail associated $\alpha\beta$ heterodimers.  At the tip of MT, PFs curl outward from the MT cylinder surface. The tip might be in shrinking GDP-cap state or growing GTP-cap state. In contrast to the tip in shrinking GDP state, the growing GTP tip is fairly straight. Or in other words, in GTP-cap state, the angle between the curled out segment of PFs and MT surface is less than that in GDP state. In this study, we will only consider the growth and shortening of one single PF, and assume that each step of growth and shortening of one PF contributes to $L$ (nm) of the growth and shortening of the whole MT. Intuitively, $L=L_1/n$ with $L_1$ the length of one $\alpha\beta$ hetrodimer. In the numerical calculations, $L=8$ nm$/13\approx$0.615 nm is used \cite{Hill1985, Kolomeisky2001}.

\begin{figure}
  \includegraphics[width=220pt]{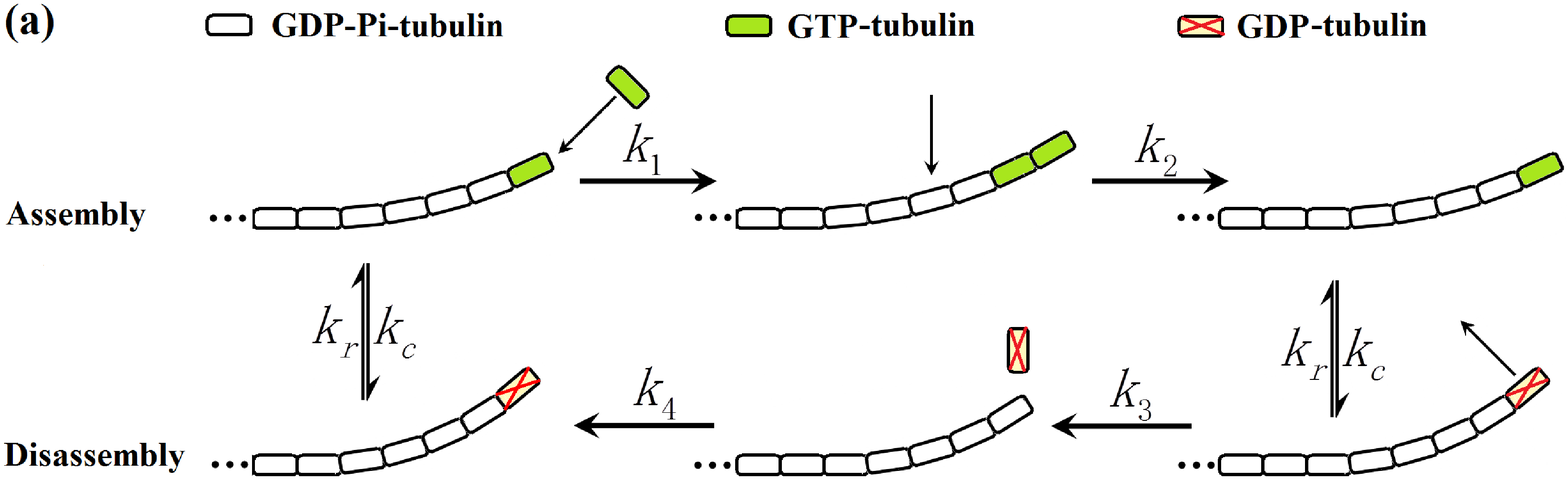}\\
  \includegraphics[width=220pt]{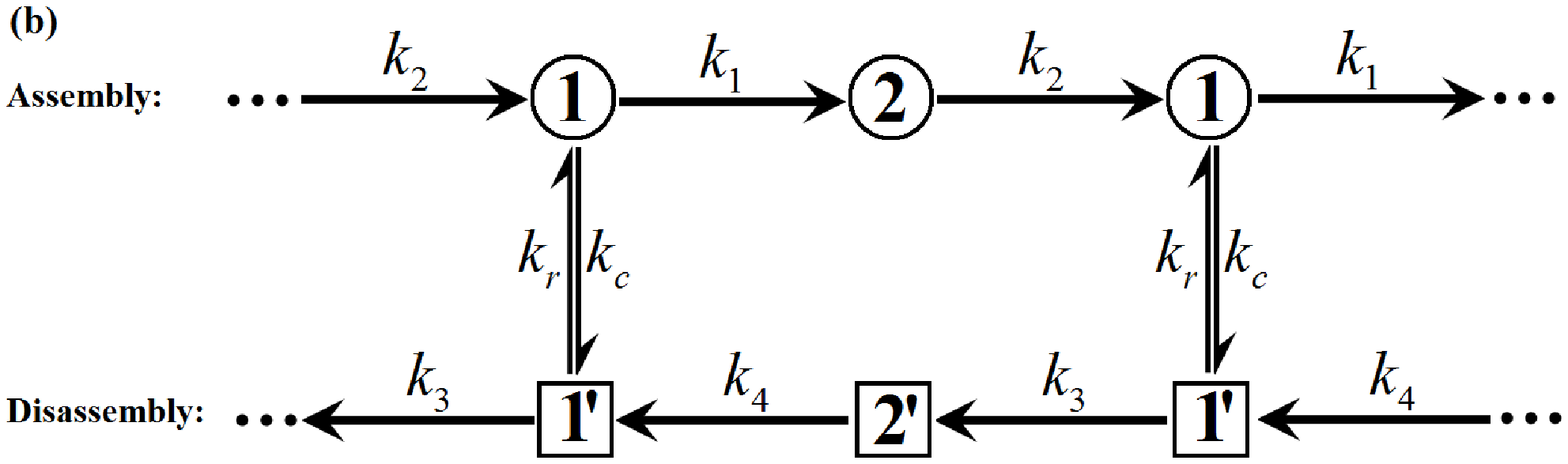}\\
  \caption{Schematic depiction of the two-state mechanochemical model of protofilament (PF) growth and shortening (a) and its corresponding two-line Markov chain (b). In assembly state, the growth of PF accomplished by two processes, one GTP-tubulin binds to the tip of PF (with GTP-tubulin concentration dependent rate $k_1$) and one PF unit rearranges into the MT surface (with external force dependent rate $k_2$). The energy used in the second process comes from the GTP hydrolysis in the penultimate tubulin unit. One tubulin unit binding to the tip of PF is assumed to be equivalent to $L$ (nm) growth of the whole MT ($L=0.615$ nm is used in this study \cite{Hill1985, Kolomeisky2001}). Similarly, in disassembly state, the shortening of PF also includes two processes, one PF unit detaches from the tip of PF and one new PF unit curls out the MT surface. In this depiction, the same as in \cite{Asbury2011}, a segment 5 dimers in length is assumed to curls out from the MT surface. }\label{FigSchematicAll}
\end{figure}

Our two-state mechanochemical model for PF growth and shortening is schematically depicted in Fig. \ref{FigSchematicAll}(a), and mathematically described by a two-line Markov chain in Fig. \ref{FigSchematicAll}(b). In this model, PF stochastically switches between two states: assembly state and disassembly state. During assembly state, PF grows through two processes, {\bf (i)} $\bf 1\rightarrow2$: free GTP-tubulin binding process with GTP-tubulin concentration (denoted by [Tubulin]) dependent rate constant $k_1$, and {\bf (ii)} $\bf 2\rightarrow1$: PF conformation change process, during which, using energy released from GTP hydrolysis, the curled PF segment is straightened with one PF unit (i.e. one $\alpha\beta$ heterodimer) rearranged into the MT surface, i.e. to parallel the MT axis approximately. During disassembly state, each step of PF shortening includes also two processes, {\bf (i')} $\bf 2'\leftarrow1'$: disassociation of GDP-tubulin from PF tip to environment and {\bf (ii')} $\bf 1'\leftarrow2'$: one new PF unit curls out from the MT surface (during which phosphate is released from the tip tubulin unit simultaneously).

The two-state model presented here can be regarded as a generalization of the one employed by B. Akiyoshi {\it et al} to explain their experimental data \cite{Akiyoshi2010}, which can be depicted by Fig. \ref{FigschematicDetachment}(a) [our corresponding generalized two-state model including bead detachment from MT is depicted in Fig. \ref{FigschematicDetachment}(b), see Sec. II.A for detailed discussion]. The reasons that we prefer to use this generalized model are that, the simple model of B. Akiyoshi {\it et al} cannot fit the measured attachment lifetime of bead on MT well (see Fig. 4(a) in \cite{Akiyoshi2010}), and moreover, the measurements in \cite{Walker1988, Walker1991, Janson2003} indicate that the rate of catastrophe, i.e. transition from elongation to shortening, dependent on GTP-tubulin concentration of the solution. However, for the simple model depicted in Fig. \ref{FigschematicDetachment}(a), the catastrophe rate $k_c$ is independent of GTP-tubulin concentration (it is biochemically reasonable to assume that the elongation rate $k_1$ depend on GTP-tubulin concentration, $k_1=k_1^0$[Tubulin], but with no reasons to write $k_c$ as as a function of [Tubulin]). We will see from the Results section that, for our generalized model, the catastrophe rate does change with [Tubulin], since GTP-tubulin concentration will change the probabilities of PF in states 1 and 2, and consequently change the transition rate from assembly state to disassembly state. At the same time, for the simple model depicted in Fig. \ref{FigschematicDetachment}(a), the distribution of catastrophe time is an exponential. However, the experimental measurement under a particular situation indicates this distribution is clearly not an exponential \cite{Janson2003} \footnote{From the parameter values listed in Tab. \ref{table1}, one can see that the rates $k_1^0$ and $k_3$ are much larger than $k_2^0$ and $k_4^0$, so under low external force and high free GTP-tubulin concentration, the model depicted in Fig. \ref{FigschematicDetachment}(a) is a good approximation of our generalized model depicted in \ref{FigschematicDetachment}(b).}.
It should be pointed out, although our model presented here looks more complex, there are only two more parameters than the one depicted in Fig. \ref{FigschematicDetachment}(a) \footnote{To keep as less parameters as possible, in our two-state model, we assume that, the bead only can detach from MT from sub-states 1 and $1'$. The reasons are as follows: in assembly state, the experimental data in \cite{Dogterom1997, Akiyoshi2010} [or see Fig. \ref{FigFittingData}(b)] imply the growth speed of MT increases with external force (Note, the definition of force direction in \cite{Akiyoshi2010} is different from that in \cite{Dogterom1997}. In this study, the force direction definition is the same as in \cite{Akiyoshi2010}, i.e., the force is positive if it points to the MT growth direction), so the corresponding force distribution factor $\delta_g$ [see Eq. (\ref{eq15})] should be positive since the growth speed $V_g=k_1k_2L/(k_1+k_2)$ [see Eq. (\ref{eq5})]. Consequently, the probability $\bar p_1={k_2}/{(k_1+k_2)}$ that MT in state 1 [see Eq. (\ref{eq12})] increases but the probability $\bar p_2={k_1}/{(k_1+k_2)}$ that MT in state 2 decreases with external force, i.e., as the increase of external force, the assembly MT would more like to stay in state 1. Meanwhile, from the experimental data in \cite{Akiyoshi2010} one sees the detachment rate from assembly state increases with external force. Therefore, the more reasonable choice is to assume that the bead can only detach from state 1 but not state 2. Through similar discussion, one also can see that it is more reasonable to assume that, in disassembly state, the bead can only detach from state $1'$. At the same time, the experimental data in \cite{Walker1988} imply the catastrophe rate decreases with GTP-tubulin concentration [Tubulin] [or see Fig. \ref{FigFittingDataExtra}(b)]. Since $k_1=k_1^0$[Tubulin], the probability $\bar p_1={k_2}/{(k_1+k_2)}$ decreases with [Tubulin], but the probability $\bar p_2={k_1}/{(k_1+k_2)}$ increases with [Tubulin]. This is why we assume the catastrophe takes place at state 1.}.

The organization of this paper is as follows. The two-state mechanochemical model will be presented and theoretically studied in the next section, and then in Sec. III, based on the model parameters obtained by fitting the experimental data mainly obtained in \cite{Akiyoshi2010}, properties of MT growth and shortening are numerically studied, including its external force and GTP-tubulin concentration dependent growth and shortening speeds, mean dwell times in assembly and disassembly state, mean growth or shortening length before the bead, used in experiments to apply external force, detachment from MT.
To know more properties about the MT dynamics, eleven {\it critical forces} (detailed definitions will be given in Sec. III) are also numerically discussed in Sec. III. Finally, Sec IV includes conclusions and remarks.

\section{Two-state mechanochemical model of protofilament}
As the schematic depiction in Fig. \ref{FigSchematicAll}, PF might be in two states, assembly (growth) state and disassembly (shortening) state. Each of the two states includes two sub-states, denoted by 1, 2 and $1'$, $2'$ respectively.
Let $p_1, p_2$ be the probabilities that PF in assembly sub-states 1 and 2 respectively, and $\rho_1, \rho_2$ be the probabilities that the PF in disassembly sub-states $1'$ and $2'$, then $p_1, p_2, \rho_1, \rho_2$ are governed by the following master equation 
\begin{equation}\label{eq1}
\begin{aligned}
dp_1/dt=&k_2p_2-k_1p_1+k_r\rho_1-k_cp_1,\cr
dp_2/dt=&k_1p_1-k_2p_2,\cr
d\rho_1/dt=&k_4\rho_2-k_3\rho_1-k_r\rho_1+k_cp_1,\cr
d\rho_2/dt=&k_3\rho_1-k_4\rho_2.
\end{aligned}
\end{equation}
Where $k_1$ is the rate of GTP-tubulin binding to the tip of PF, $k_2$ is the rate of PF realignment with one new unit lying into the MT surface, $k_3$ is the dissociation rate of GDP-tubulin from the tip of PF, and $k_4$ is the rate of curling out of one tubulin unit from the MT surface (with Pi release simultaneously). The steady state solution of Eq. (\ref{eq1}) is
\begin{equation}\label{eq2}
\begin{aligned}
&p_1=\left[1+\frac{k_1}{k_2}+\frac{k_c}{k_r}\left(1+\frac{k_3}{k_4}\right)\right]^{-1},\cr
&p_2=\frac{k_1}{k_2}p_1,\quad \rho_1=\frac{k_c}{k_r}p_1,\quad
\rho_2=\frac{k_3}{k_4}\frac{k_c}{k_r}p_1.
\end{aligned}
\end{equation}
One can easily show that the mean steady state velocity of MT growth or shortening is \cite{Derrida1983, Zhang20092}
\begin{equation}\label{eq3}
\begin{aligned}
V=(k_2p_2-k_4\rho_2)L=\left(k_1-k_3{k_c}/{k_r}\right)p_1L,
\end{aligned}
\end{equation}
where $L$ is the effective step size of MT growth corresponding to one step growth of one PF, and $V<0$ means MT is shortening in long time average with speed $-V$.

Let $\bar p_1, \bar p_2$ be the probabilities that PF in sub-state 1 and sub-state 2 respectively, provided the PF is in assembly state, then $\bar p_1, \bar p_2$ satisfy
\begin{equation}\label{eq4}
\begin{aligned}
d\bar p_1/dt=&k_2\bar p_2-k_1\bar p_1=-d\bar p_2/dt.
\end{aligned}
\end{equation}
One can easily get that, at steady state, the mean growth speed of MT with a PF in assembly state is
\begin{equation}\label{eq5}
\begin{aligned}
V_g=k_2\bar p_2L=\frac{k_1k_2L}{k_1+k_2}.
\end{aligned}
\end{equation}
Similarly, the mean shortening speed of MT with a PF in disassembly state is
\begin{equation}\label{eq6}
\begin{aligned}
V_s=\frac{k_3k_4L}{k_3+k_4}.
\end{aligned}
\end{equation}

\subsection{Modified model according to experiments: including bead   detachment from MT}
\begin{figure}
  \includegraphics[width=180pt]{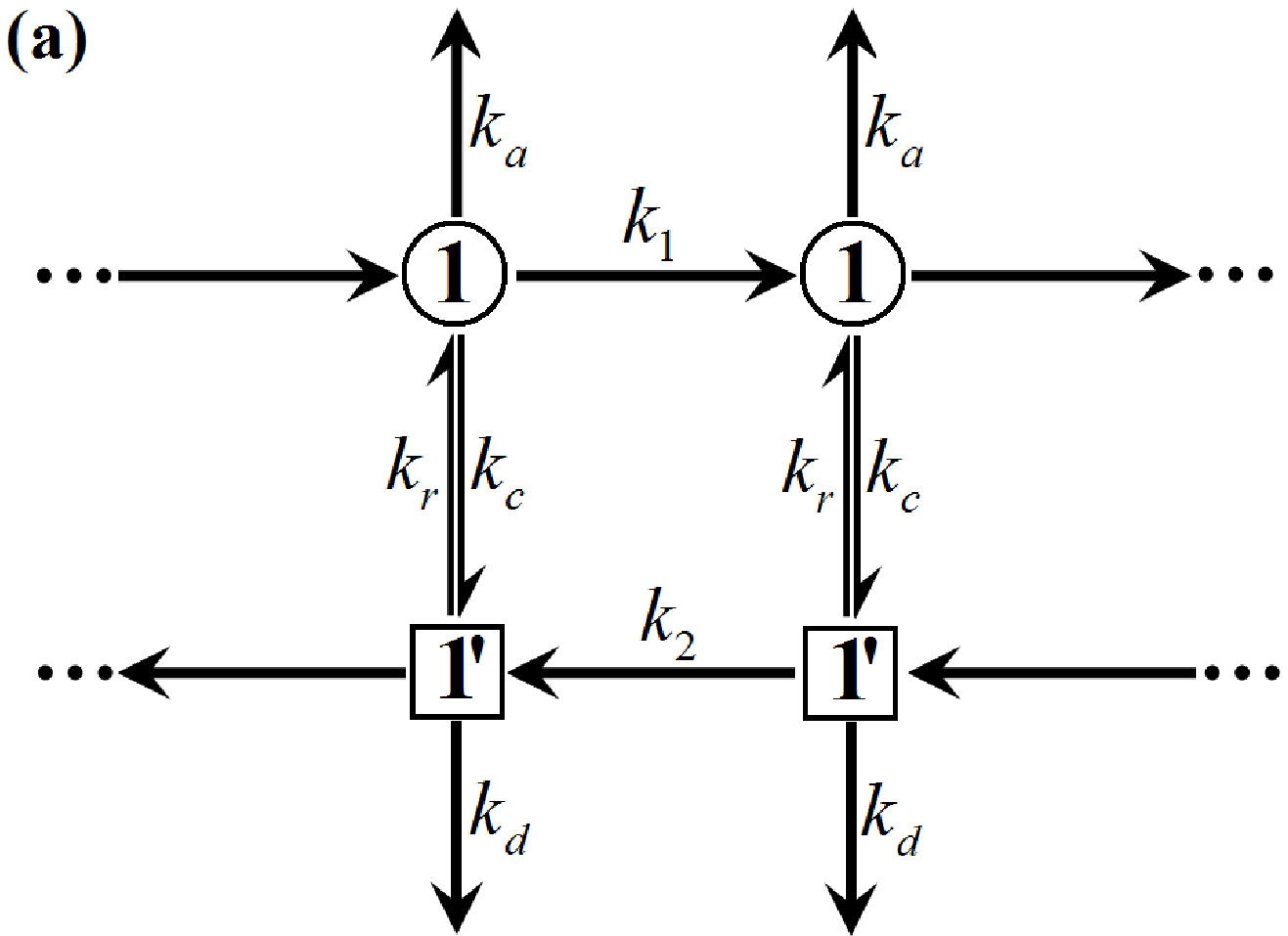}\\
  \includegraphics[width=220pt]{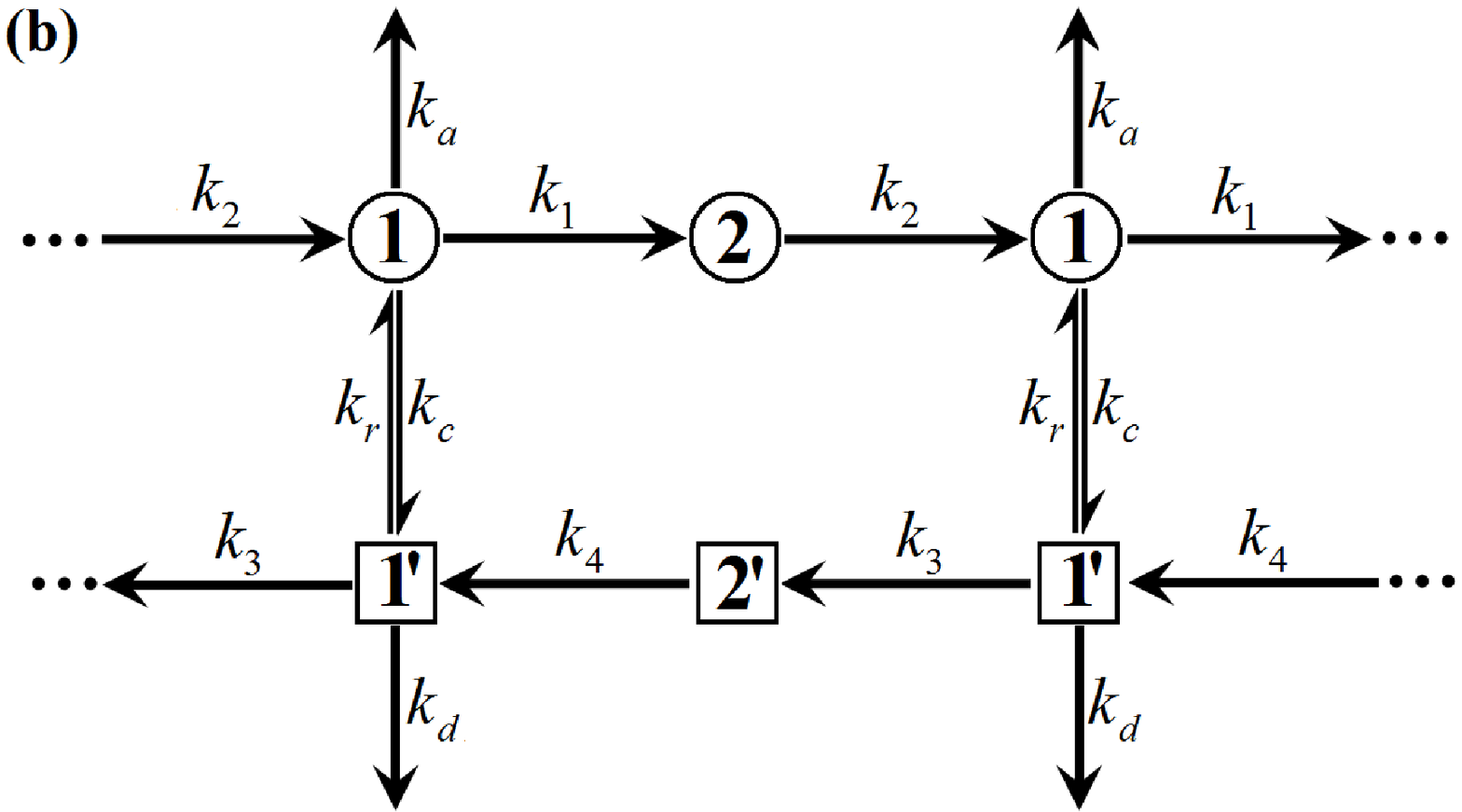}\\
  \caption{(a) Schematic depiction of the two-state model used by  B. Akiyoshi {\it et al} in \cite{Akiyoshi2010}.  In which, both the assembly and disassembly of PF are assumed to include only one process, described by rates $k_1$ and $k_2$ respectively. (b) Modified mechanochemical model with bead detachment. In the experiments of \cite{Akiyoshi2010}, Akiyoshi {\it et al} attached a bead prepared with kinetochore to the growing end of MTs, and measured not only the force dependent mean growing and shortening speeds, switch rates between assembly and disassembly states (i.e. rates of rescue and catastrophe), but also the mean lifetime of the bead on MTs, and the rates of bead detachment during assembly and disassembly states respectively. Therefore, to get the model parameters and know more properties of MT growth and shortening, this modified model is used in this study
  .  The main difference between these two models is that, the rate of detachment from assembly state and the rate of catastrophe in model (b) depend on GTP-tubulin concentration [Tubulin], but they do not in model (a). }\label{FigschematicDetachment}
\end{figure}

To know the model parameters $k_i$, $i=1,\cdots, 4$ and $k_c, k_r$, we need to fit the model with experimental data. In recent experiments \cite{Akiyoshi2010}, Akiyoshi {\it et al} attached a bead prepared with kinetochore particles to the growing end of MTs, and constant tension was applied to bead using a servo-controlled laser trap. In their experiments, not only the force dependent mean growth and shortening speeds of MTs, the rates of rescue and catastrophe, but also the force dependent mean lifetime, during which the bead is keeping attachment to MT, and mean detachment rates of the bead during assembly and disassembly states are measured. Therefore, to fit these experimental data, the model depicted in Fig. \ref{FigSchematicAll} should be modified to include the bead detachment processes [see Fig. \ref{FigschematicDetachment}(b)].

For the model depicted in Fig. \ref{FigschematicDetachment}(b), the formulations of mean growth velocity $V$, mean growth and shortening speeds $V_g$ and $V_s$ are the same as in Eqs. (\ref{eq3}) and (\ref{eq5}) (\ref{eq6}).
In the following, we will get the expression of mean lifetime of the bead on MTs.
Let $T_{1}, T_{2}, T_{1'}, T_{2'}$ be the mean first passage times (MFPTs) of a bead to detachment with initial sub-states 1, 2, $1'$ and $2'$ respectively, then $T_{1}, T_{2}, T_{1'}, T_{2'}$ satisfy \cite{Redner2001, Pury2003, Kolomeisky2005}
\begin{equation}\label{eq7}
\begin{aligned}
T_{1}=&\frac{1}{k_1+k_{a}+k_c}+\frac{k_1}{k_1+k_{a}+k_c}T_{2}
+\frac{k_c}{k_1+k_{a}+k_c}T_{1'},\cr
T_{1'}=&\frac{1}{k_r+k_3+k_{d}}+\frac{k_3}{k_r+k_3+k_{d}}T_{2'}+
\frac{k_r}{k_r+k_3+k_{d}}T_{1},\cr
T_{2}=&\frac{1}{k_2}+T_{1},\quad
T_{2'}=\frac{1}{k_4}+T_{1'}.
\end{aligned}
\end{equation}
Then the mean lifetime can be obtained as follows
\begin{equation}\label{eq8}
\begin{aligned}
T=p_1T_1+p_2T_2+\rho_1T_{1'}+\rho_2T_{2'},
\end{aligned}
\end{equation}
where $p_1, p_2, \rho_1, \rho_2$ can be obtained by formulations in Eq. (\ref{eq2}).

In assembly state, let $T_{a1}$ and $T_{a2}$ be the MFPTs to detachment of the bead initially at sub-states 1 and 2 respectively, then $T_{a1}, T_{a2}$ satisfy
\begin{equation}\label{eq9}
\begin{aligned}
T_{a1}=\frac{1}{k_1+k_{a}}+\frac{k_1}{k_1+k_{a}}T_{a2},\quad
T_{a2}=\frac{1}{k_2}+T_{a1}.
\end{aligned}
\end{equation}
One can easily show
\begin{equation}\label{eq10}
\begin{aligned}
T_{a1}=\frac{k_1+k_2}{k_2k_{a}}, \quad T_{a2}=\frac{k_1+k_{a}+k_2}{k_2k_{a}}.
\end{aligned}
\end{equation}
Therefore, the MFPT to detachment of the bead in assembly state is
\begin{equation}\label{eq11}
\begin{aligned}
T_{a}=\bar p_1T_{a1}+\bar p_2T_{a2},
\end{aligned}
\end{equation}
where the steady state probabilities
\begin{equation}\label{eq12}
\begin{aligned}
\bar p_1=\frac{k_2}{k_1+k_2},\quad \bar p_2=\frac{k_1}{k_1+k_2},
\end{aligned}
\end{equation}
are obtained from Eq. (\ref{eq4}).
The mean detachment rate during assembly can then be obtained by $K_a=1/T_{a}=1/(\bar p_1T_{a1}+\bar p_2T_{a2})$, i.e.,
\begin{equation}\label{eq13}
\begin{aligned}
K_a=\frac{(k_1+k_2)k_2k_{a}}{(k_1+k_2)^2+
k_1k_{a}}.
\end{aligned}
\end{equation}
Similarly, the mean detachment rate during disassembly can be obtained as follows
\begin{equation}\label{eq14}
\begin{aligned}
K_d=&1/T_d=1/(\bar \rho_1T_{d1}+\bar \rho_2T_{d2})\cr
=&\frac{(k_3+k_4)k_4k_{d}}{(k_3+k_4)^2+
k_3k_{d}},
\end{aligned}
\end{equation}
with steady state probabilities
$\bar \rho_1=k_4/(k_3+k_4),\ \bar \rho_2=k_3/(k_3+k_4)$.

Let $T_{c1}$ and $T_{c2}$ be the MFPTs of MT to catastrophe from sub-states 1 and 2 respectively, then $T_{c1}$ and $T_{c2}$ satisfy [see Fig. \ref{FigschematicDetachment}(b)]
\begin{equation}\label{eq141}
\begin{aligned}
T_{c1}=\frac{1}{k_1+k_c}+\frac{k_1}{k_1+k_c}T_{c2},\quad
T_{c2}=\frac{1}{k_2}+T_{c1}.
\end{aligned}
\end{equation}
The mean rate of catastrophe can be obtained by $K_c=1/T_c=1/(\bar p_1T_{c1}+\bar p_2T_{c2})$. The explicit expression can be obtained by replace $k_a$ with $k_c$ in Eq. (\ref{eq13})
\begin{equation}\label{eq142}
\begin{aligned}
K_c=\frac{(k_1+k_2)k_2k_{c}}{(k_1+k_2)^2+
k_1k_{c}}.
\end{aligned}
\end{equation}
Similarly, the mean rate of rescue is
\begin{equation}\label{eq143}
\begin{aligned}
K_r=\frac{(k_3+k_4)k_4k_{r}}{(k_3+k_4)^2+k_3k_{r}}.
\end{aligned}
\end{equation}

\subsection{Force and GTP-tubulin concentration dependence of the transition rates}

From the experimental data in \cite{Akiyoshi2010} [or see Fig. \ref{FigFittingData}], one sees some transition rates in our model should depend on the external force. Since the processes $1\rightarrow2$ and $2'\leftarrow1'$ are accomplished by binding tubulin unit to and releasing tubulin unit from the tip of PF [see Fig. \ref{FigSchematicAll} and \ref{FigschematicDetachment}(b)], we assume that $k_1$ and $k_3$ are force independent. Similar as the methods demonstrated in the models of molecular motors \cite{Fisher2001, Zhang20093} and models for adhesive of cells to cells \cite{Bell1978}, the external force $F$ dependence of rates $k_2$, $k_4$, $k_{a}$, $k_{d}$, $k_r$, $k_c$ are assumed to be the following forms
\begin{equation}\label{eq15}
\begin{aligned}
k_{l}=k_l^0e^{FL\delta_l/k_BT},\quad l=2, 4, a, d, r, c.
\end{aligned}
\end{equation}
Hereafter, the external froce $F$ is positive if it points to the direction of MT growth.

Meanwhile, the rate $k_1$ should depend on the concentration of free GTP-tubulin in solution. Similar as the method in \cite{Fisher2001}, we simply assume $k_1=k_1^0$[Tubulin].

\subsection{Critical forces of MT growth}
For the sake of the better understanding of external force $F$ dependent properties of MTs and the experimental verification of the two-state model, in the following, we will define altogether eleven {\it critical forces}. Corresponding numerical results will be presented in the next section.

\noindent {\bf (1) }{\it Critical Force $F_{c1}$}: under which $V_g(F_{c1})=V_s(F_{c1})$, i.e. the average speeds of assembly and disassembly are the same. From Eqs. (\ref{eq5}) (\ref{eq6}) one sees $F_{c1}$ satisfies
\begin{equation}\label{eq16}
\begin{aligned}
k_1k_2(F_{c1})[k_3+k_4(F_{c1})]=k_3k_4(F_{c1})[k_1+k_2(F_{c1})].
\end{aligned}
\end{equation}

\noindent {\bf (2) }{\it Critical Force $F_{c2}$}: under which the mean velocity of MT growth is vanished. Formulation (\ref{eq3}) gives $k_1k_r(F_{c2})=k_3k_c(F_{c2})$, i.e.
\begin{equation}\label{eq17}
\begin{aligned}
F_{c2}=&\frac{k_BT}{(\delta_r-\delta_c)L}\ln \frac{k_3k_c^0}{k_1k_r^0}\cr
=&\frac{k_BT}{(\delta_r-\delta_c)L}\ln \frac{k_3k_c^0}{k_1^0k_r^0\textrm{[Tubulin]}}.
\end{aligned}
\end{equation}

\noindent {\bf (3) }{\it Critical Force $F_{c3}$}: under which $p_1+p_2=\rho_1+\rho_2$, i.e., the probabilities that MTs in assembly and disassembly states are the same. From expressions in Eq. (\ref{eq2}), one easily sees $F_{c3}$ satisfies
\begin{equation}\label{eq18}
\begin{aligned}
&k_r(F_{c3})k_4(F_{c3})[k_1+k_2(F_{c3})]\cr
=&k_c(F_{c3})k_2(F_{c3})[k_3+k_4(F_{c3})].
\end{aligned}
\end{equation}

\noindent {\bf (4) }{\it Critical Force $F_{c4}$}: under which the detachment rates during assembly and disassembly states are the same. In view of formulations (\ref{eq13}) and (\ref{eq14}), one can get $F_{c4}$ by $K_a(F_{c4})=K_d(F_{c4})$.

\noindent {\bf (5) }{\it Critical Force $F_{c5}$}: under which the mean dwell times of MT in assembly and disassembly states are the same.

Let $T_{g1}$ and $T_{g2}$ be the MFPTs of bead to detachment or catastrophe of MT with initial sub-states 1 and 2 respectively, then $T_{g1}, T_{g2}$ satisfy (see Fig. \ref{FigschematicDetachment}(b) and Refs. \cite{Redner2001, Pury2003})
\begin{equation}\label{eq19}
\begin{aligned}
T_{g1}=&\frac{1}{k_1+k_{a}+k_c}+\frac{k_1}{k_1+k_{a}+k_c}T_{g2},\cr
T_{g2}=&\frac{1}{k_2}+T_{g1}.
\end{aligned}
\end{equation}
Its solution is
\begin{equation}\label{eq20}
\begin{aligned}
T_{g1}=\frac{k_1+k_2}{k_2(k_{a}+k_c)},\quad
T_{g2}=\frac{k_1+k_{a}+k_c+k_2}{k_2(k_{a}+k_c)}.
\end{aligned}
\end{equation}
The mean dwell time of MT in assembly (or growth) state is then
\begin{equation}\label{eq21}
\begin{aligned}
T_{g}=\bar p_1T_{g1}+\bar p_2T_{g2}=\frac{(k_1+k_2)^2+k_1(k_a+k_c)}{k_2(k_1+k_2)(k_a+k_c)}.
\end{aligned}
\end{equation}
Similarly, the mean dwell time of MT in disassembly (or shortening) state can be obtained as follows
\begin{equation}\label{eq22}
\begin{aligned}
T_{s}=\frac{(k_3+k_4)^2+k_3(k_d+k_r)}{k_4(k_3+k_4)(k_d+k_r)}.
\end{aligned}
\end{equation}
The critical force $F_{c5}$ can then be obtained by $T_{g}(F_{c5})=T_{s}(F_{c5})$.

\noindent {\bf (6) }{\it Critical Force $F_{c6}$}: under which the mean lifetime of the bead on MT attains its maximum, i.e. $T(F_{c6})=\max_{F}T(F)$ with $T$ given by formulation (\ref{eq8}).

\noindent {\bf (7) }{\it Critical Force $F_{c7}$}: under which the mean growth length of MT attains its maximum.
The mean growth length of MT can be obtained by $l_+=VT$ with $V, T$ satisfy formulations (\ref{eq3}) and (\ref{eq8}) respectively.

\noindent {\bf (8) }{\it Critical Force $F_{c8}$}: under which the mean shortening length of MT attains its maximum. The mean shortening length of MT can be obtained by $l_-=-VT$ with $V, T$ satisfy formulations (\ref{eq3}) and (\ref{eq8}) respectively.

\noindent {\bf (9) }{\it Critical Force $F_{c9}$}: the rates of catastrophe and rescue are the same, i.e. $K_c(F_{c9})=K_r(F_{c9})$ [see Eqs. (\ref{eq142}) and (\ref{eq143})]. Under {\it critical force} $F_{c9}$, the average switch time between growth and shortening, i.e. $1/K_c$ and $1/K_r$, are the same. It is to say that the mean duration for each growth and each shortening period are the same.

\noindent {\bf (10) }{\it Critical Force $F_{c10}$}: under which $V_gT_g=V_sT_s$. Here $V_gT_g=:l_g$ is the mean growth length before bead detachment or catastrophe, and $V_sT_s=:l_s$ is the mean shortening length before bead detachment or rescue. The formulations of $V_g, V_s$ and $T_g, T_s$ are in Eqs. (\ref{eq5}) (\ref{eq6}) and (\ref{eq21}) (\ref{eq22}).

\noindent {\bf (11) }{\it Critical Force $F_{c11}$}: under which $V_g/K_c=V_s/K_r$. Here $V_g/K_c=:l_g^*$ is the mean growth length before catastrophe, and $V_s/K_r=l_s^*$ is the mean shortening length before rescue. The formulations of $K_c, K_r$ are in Eqs. (\ref{eq142}) (\ref{eq143}).

It needs to be clarified that, the definitions for $F_{c1}, F_{c2}, F_{c3}, F_{c9}, F_{c11}$ are unrelated to bead detachment, but the definitions for $F_{c4}, F_{c5}, F_{c6}, F_{c7}, F_{c8}, F_{c10}$ do. Therefore the values of $F_{c1}, F_{c2}, F_{c3}, F_{c9}, F_{c11}$ obtained in this theoretical study can be verified by various experimental methods as in \cite{Walker1988, Walker1991, Verde1992, Dogterom1997, Janson2003, Grishchuk2005, Joglekar2010}, but the values of $F_{c4}, F_{c5}, F_{c6}, F_{c7}, F_{c8}, F_{c10}$ can only be verified by similar experimental method as in \cite{Akiyoshi2010}.
For the sake of convenience, and based on the above definitions and numerical calculations in Sec. III (see Figs. \ref{FigCriticalForce} and \ref{FigCriticalForceDetail}), basic properties of the eleven {\it critical forces} $F_{ci}$ are listed in Tab. \ref{table2}. Meanwhile, the main symbols used in this study are listed in Tab. \ref{table3}.
\begin{table}
  \centering
  \caption{Basic properties of the {\it critical forces} as defined in Sec. II.C, see also Figs. \ref{FigCriticalForce} and \ref{FigCriticalForceDetail}.}
    \begin{tabular}{c|c|c|c}
    \hline\hline
      $i$ & $F<F_{ci}$ & $F=F_{ci}$  & $F>F_{ci}$ \\
    \hline
     1 & $V_g<V_s$ & $V_g=V_s$ & $V_g>V_s$\\
     2 & $V<0$ & $V=0$ & $V>0$\\
     3 & $p_1+p_2<\rho_1+\rho_2$ & $p_1+p_2=\rho_1+\rho_2$ & $p_1+p_2>\rho_1+\rho_2$\\
     4 & $K_a<K_d$ & $K_a=K_d$ & $K_a>K_d$\\
     5 & $T_g>T_s$ & $T_g=T_s$ & $T_g<T_s$\\
     6 & $T<\max_FT$ & $T=\max_FT$ & $T<\max_FT$\\
     7 & $l_+<\max_Fl_+$ & $l_+=\max_Fl_+$ & $l_+<\max_Fl_+$\\
     8 & $l_-<\max_Fl_-$ & $l_-=\max_Fl_-$ & $l_-<\max_Fl_-$\\
     9 & $K_r<K_c$ & $K_r=K_c$ & $K_r>K_c$\\
     10 & $l_g<l_s$ & $l_g=l_s$ & $l_g>l_s$\\
     11 & $l_g^*<l_s^*$ & $l_g^*=l_s^*$ & $l_g^*>l_s^*$\\
    \hline\hline
  \end{tabular}
  \label{table2}
\end{table}
\begin{table}
  \centering
  \caption{The main symbols and their expressions (or definitions) used in this study.}
    \begin{tabular}{c|c|c}
    \hline\hline
      Symbol & Biophysical meaning & Definitions \\
    \hline
     $V$ & mean velocity of MTs & Eq. (\ref{eq3}) \\
     $V_g$ & growth speed of MTs & Eq. (\ref{eq5}) \\
     $V_s$ & shortening speed of MTs & Eq. (\ref{eq6}) \\
     $T$ & mean lifetime of bead & Eq. (\ref{eq8}) \\
     $K_a$ & bead detachment rate (assembly) & Eq. (\ref{eq13}) \\
     $K_d$ & bead detachment rate (disassembly) & Eq. (\ref{eq14}) \\
     $T_a$ & time to detachment (assembly )& $1/K_a$ \\
     $T_d$ & time to detachment (assembly )& $1/K_d$ \\
     $K_c$ & catastrophe rate & Eq. (\ref{eq142}) \\
     $K_r$ & rescue rate & Eq. (\ref{eq143}) \\
     $p_i, \rho_i$ & probability & Eq. (\ref{eq2}) \\
     $T_g$ & mean growth time& Eq. (\ref{eq21}) \\
     $T_s$ & mean shortening time& Eq. (\ref{eq22}) \\
     $l_g, l_s$ & $l_g=V_gT_g,\ l_s=V_sT_s$& see $F_{c10}$ \\
     $l_g^*, l_s^*$ & $l_g^*=V_g/K_c,\ l_s^*=V_s/K_r$& see $F_{c11}$ \\
     $l_+, l_-$ & $l_+=VT,\ l_-=-VT$& see $F_{c7,8}$ \\
     $k_l$ & rate constants [Fig. \ref{FigschematicDetachment}(2)] & Eq. (\ref{eq15}) \\
    \hline\hline
  \end{tabular}
  \label{table3}
\end{table}

\section{Results}
\begin{figure}
  \includegraphics[width=270pt]{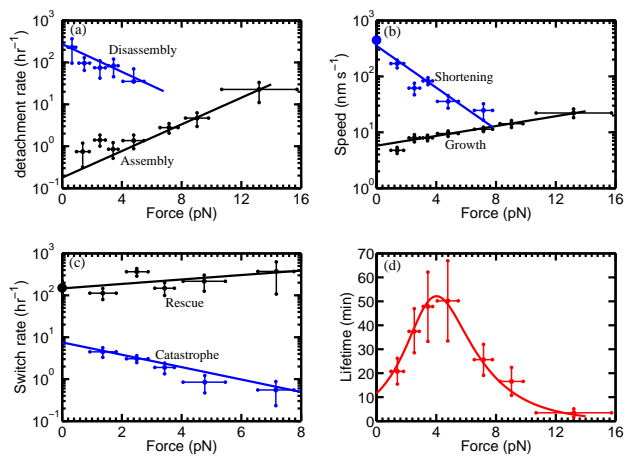}\\
  \caption{Theoretical results of the two-state model [see Fig. \ref{FigschematicDetachment}(b)] and experimental data obtained by Akiyoshi {\it et al} \cite{Akiyoshi2010}: The detachment rates are obtained by formulations (\ref{eq13}) (\ref{eq14}), the speeds are obtained by formulations (\ref{eq5}) (\ref{eq6}), the switch rates are obtained by formulations (\ref{eq142}) (\ref{eq143}), and the lifetime is obtained by formulation (\ref{eq8}). The model parameters used in the theoretical calculations are listed in Tab. \ref{table1}. The two black dots on vertical axis of (b) and (c) are obtained by averaging the data in \cite{Walker1988} [see the lines in Fig. \ref{FigFittingDataExtra}(a) and Fig. \ref{FigFittingDataExtra}(b)].}\label{FigFittingData}
\end{figure}

\begin{figure}
  \includegraphics[width=270pt]{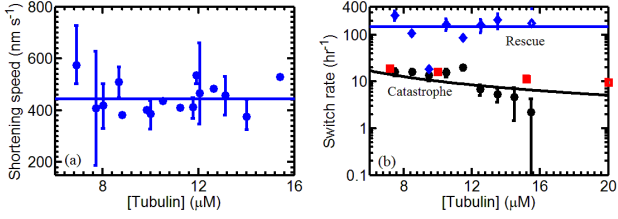}\\
  \caption{GTP-tubulin concentration dependent data measured by Walker {\it et al} \cite{Walker1988}: (a) Shortening speed of MT and their average value.  (b) Switch rates of MT between assembly and disassembly, where the curve is obtained by our theoretical model using the parameter listed in Tab. \ref{table1} (see formulation (\ref{eq142}) with $k_1=k_1^0$[Tubulin]), the solid squares are experimental data from \cite{Janson2003}. }\label{FigFittingDataExtra}
\end{figure}

In order to discuss the properties of MT growth and shortening, the model parameters, i.e. $k_1^0, k_3$ and $k_i^0$, $\delta_i$ for $i=2,4,a,d,r,c$ should be firstly obtained. By fitting the expressions of $V_g, V_s, T, K_a, K_d, K_c, K_r$, which are given in Eqs. (\ref{eq5}) (\ref{eq6}) (\ref{eq8}) (\ref{eq13}) (\ref{eq14}) (\ref{eq142}) (\ref{eq143}) respectively, to the experimental data mainly measured in \cite{Akiyoshi2010}, these parameter values are obtained (see Fig. \ref{FigFittingData} and Tab. \ref{table1}, the fitting methods are discussed in \footnote{In our fitting, we firstly get the parameters $k_1^0, k_2^0, \delta_g$ and $k_3, k_4^0, \delta_s$ by fitting formulations (\ref{eq5}) and (\ref{eq6}) to the experimental data of growth and shortening speeds respectively [see Fig. \ref{FigFittingData}(b)], and then get $k_r^0, \delta_r$ and $k_c^0, \delta_c$ by fitting formulations (\ref{eq142}) and (\ref{eq143}) to the catastrophe and rescue rates [see Fig. \ref{FigFittingData}(c)], $k_a^0, \delta_a$ and $k_d^0, \delta_d$ are determined by fitting formulations (\ref{eq13}) and (\ref{eq14}) to the corresponding data plotted in Fig. \ref{FigFittingData}(a). Finally all the parameters are slightly adjusted according to the experimental data about the mean lifetime of bead attachment to MT [see formulation (\ref{eq8}) and Fig. \ref{FigFittingData}(d)]. All the fitting are done by the nonlinear least square program {\it lsqnonlin} in Matlab. In each fitting, We randomly choose 1000 initial values of the parameters and adopt the parameter values which fit the experimental data best.}). The data corresponding to zero external force in Figs \ref{FigFittingData}(b) and \ref{FigFittingData}(c) [the two black dots on vertical axis] are obtained by fitting the corresponding measurement in \cite{Walker1988} with a constant [see the two lines in  Figs. \ref{FigFittingDataExtra}(a) and \ref{FigFittingDataExtra}(b)], since as implied by our model, the rates of MT shortening and rescue are independent of GTP-tubulin concentration.  All the following calculations will be based on the parameters listed in Tab. \ref{table1}. The curve in Fig. \ref{FigFittingDataExtra}(b) is the theoretical prediction of GTP-tubulin concentration dependent catastrophe rate $K_c$ by formulation (\ref{eq142}). Compared with the experimental data measured in \cite{Walker1988, Janson2003}, these prediction looks satisfactory \footnote{The parameter values listed in Tab. \ref{table1} do not fit well to the GTP-tubulin concentration dependent growth speed of MTs obtained in \cite{Walker1988, Janson2003}, since the corresponding data in \cite{Walker1988, Janson2003} are much different from that in \cite{Akiyoshi2010}. Without external force, but under similar GTP and tubulin concentration, the growth speed of MT measured in \cite{Walker1988} is about 43 nm/s, and about 20 nm/s in \cite{Janson2003}, but it is only about 5 nm/s in \cite{Akiyoshi2010}. In this study, we get the parameter values mainly based on the data measured in \cite{Akiyoshi2010}. One reason is that, from our model, if the GTP-tubulin concentration is nonzero, the growth speed of MT will always positive [see formulation (\ref{eq5}), $V_g\gneq0$ if $k_1\gneq0$]. However, this might not be true for the data in \cite{Walker1988, Janson2003}. So, it might be impossible to get a believable fitting parameters for formulation (\ref{eq5}) from data in \cite{Walker1988, Janson2003} since the data in \cite{Walker1988, Janson2003} cannot be described by a formulation like (\ref{eq5}). In \cite{Dogterom1997}, the velocity-force data are measured under tubulin concentration 25$\mu$M. However, the zero force growth speed obtained there is about 20 nm/s, which is also much larger than that obtained in \cite{Akiyoshi2010}. Consequently, the theoretical results based on the parameter values listed in Tab. \ref{table1} do not fit well to their data either. One can verify that the velocity-force data in \cite{Dogterom1997} can be well described by formulation (\ref{eq5}) but with parameters $k_1^0=2.99$ s$^{-1}\mu$M$^{-1}$, $k_2^0=53.17$ s$^{-1}$ and $\delta_g=4.13$. The difference among these experimental data might due to the differences of experimental techniques, methods or materials.}.
\begin{table}
  \centering
  \caption{Model parameters obtained by fitting the expressions in (\ref{eq5}) (\ref{eq6}) (\ref{eq8}) (\ref{eq13}) (\ref{eq14}) (\ref{eq142}) (\ref{eq143}) to the experimental data mainly measured in \cite{Akiyoshi2010}. In the fitting, $k_BT=4.12$ pN$\cdot$nm and effective step size $L=0.615$ nm are used \cite{Hill1985, Kolomeisky2001, Joglekar2002}. The fitting results are plotted in Fig. \ref{FigFittingData}.}
    \begin{tabular}{cc||cc}
    \hline\hline
      Parameter & value  & Parameter & Value \\
    \hline
     $k_1^0$ & $5.8\times10^5$ s$^{-1}\mu$M$^{-1}$ & $k_2^0$ & 9.3 s$^{-1}$\\
     $k_3$ & $1.0\times10^8$ s$^{-1}$ & $k_4^0$ & 571.6 s$^{-1}$\\
     $k_r^0$ & $7.4\times10^3$ s$^{-1}$ & $k_c^0$ & $1.8\times10^3$ s$^{-1}$ \\
     $k_a^0$ & 41.9 s$^{-1}$& $k_d^0$ & $1.4\times10^4$ s$^{-1}$\\
     \hline
     $\delta_g$ & 0.68 & $\delta_s$ & -2.88 \\
     $\delta_r$ & 3.71 & $\delta_c$ & -2.96 \\
     $\delta_a$ & 1.77 & $\delta_d$ & 0.33\\
    \hline\hline
  \end{tabular}
  \label{table1}
\end{table}

\begin{figure}
  \includegraphics[width=270pt]{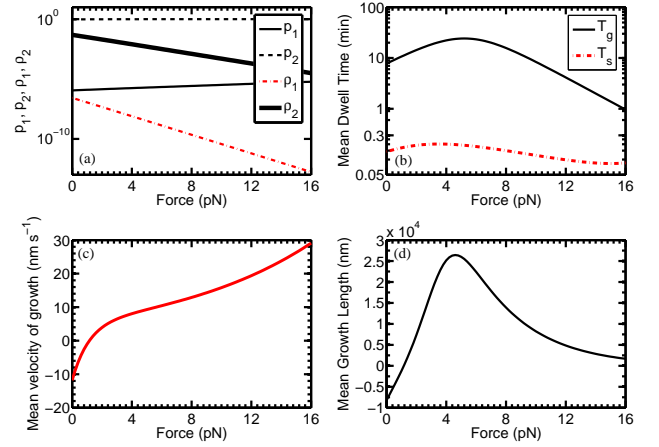}\\
  \caption{Properties of MT growth and shortening obtained by the two-state model [see Fig. \ref{FigschematicDetachment}(b)] with model parameters listed in Tab. \ref{table1}: (a) Under external force, the MT is mainly in assembly state, both the probabilities $p_1$ and $p_2$ of MT in assembly sub-states 1 and 2 increase with force but with $p_2\gg p_1$. During disassembly state, the MT is mainly in sub-state $2'$, $\rho_2>\rho_1$. Here $p_1, p_2, \rho_1, \rho_2$ are calculated by formulations in (\ref{eq2}). (b) The dwell time $T_{g}$ of MT in assembly state is always larger than that in disassembly state (denoted by $T_{s}$) for external force less than 16 pN [see formulations (\ref{eq21}) (\ref{eq22}) for $T_{g}$ and $T_{s}$]. Similar as the mean lifetime of bead attachment to MT [see Fig. \ref{FigFittingData}(d)], both $T_{g}$ and $T_{s}$ increase firstly and then decrease with external force. (c) The mean velocity of MT growth [see formulation (\ref{eq3})] increases with external force monotonically, where the negative velocity means MT shortens its length in long time average, though the curves in (a) and (b) imply that the MT is mainly in assembly state. (d) The mean growth length before bead detachment increases firstly and then deceases with external force. Here the mean growth length is obtained by mean growth velocity of MT multiplied by mean lifetime of the bead, i.e. $VT$ [see formulation (\ref{eq3}) for $V$ and formulation (\ref{eq8}) for $T$].}\label{FigProperties}
\end{figure}

\begin{figure}
  \includegraphics[width=270pt]{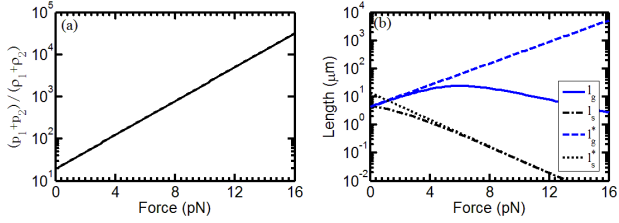}\\
  \caption{(a) The ratio of probability $p_1+p_2$ that MT in assembly state to probability $\rho_1+\rho_2$ that MT in disassembly state increases exponentially with the external force, and under positive external force, the MT mainly stays in assembly state, although the assembly speed might be much lower compared with the disassembly speed [see Fig. \ref{FigFittingData}(b)]. (b) The mean growth length $l_g, l_g^*$ and shortening length $l_s, l_s^*$ of MT in one assembly and disassembly period. The difference between $l_g, l_s$ and $l_g^*, l_s^*$ is that, in the calculation of $l_g^*, l_s^*$, the bead attached to the tip of MT, through which the external force is applied to MT, is assumed to keep attachment to MT, or the external force just applied by other methods \cite{Walker1988, Walker1991, Dogterom1997, Janson2003}, so the MT can only leave assembly state by catastrophe and leave disassembly state by rescue. }\label{FigPropertiesExtra}
\end{figure}

From Fig. \ref{FigProperties}(a), one can see that, the MT is mainly in assembly state. Further calculations indicate that the ratio of probabilities in assembly state to disassembly state, i.e. $(p_1+p_2)/(\rho_1+\rho_2)$, increases exponentially with external force $F$ [see Fig. \ref{FigPropertiesExtra}(a)]. In experiments of Akiyoshi {\it at al} \cite{Akiyoshi2010}, the external force $F$ is applied to MT through a bead attached to its growing tip. Fig. \ref{FigProperties}(b) indicates that, for $F\le 16$ pN, the mean dwell time of MT in assembly state before bead detachment is larger than that in disassembly state. Although the MT is mainly in assembly state, its mean growth velocity is negative under small external force [Fig. \ref{FigProperties}(c)], since for such cases, the shortening speed is greatly larger than the growth speed [see Fig. \ref{FigFittingData}(b)]. But, Fig. \ref{FigProperties}(c) indicates the mean velocity of MT growth always increases with external force. Similar as the mean growth velocity, the mean growth length of MT before bead detachment might be negative [i.e. MT shortens its length in long time average, see Fig. \ref{FigProperties}(d)],  though the MT spends most of its time in assembly state [Fig. \ref{FigProperties}(b)]. Similar as the mean lifetime [Fig. \ref{FigFittingData}(d)], the mean growth length of MT also has a global maximum for external force [Fig. \ref{FigProperties}(d)]. As we have mentioned in the Introduction, the chromosome segregation is accomplished by the tensile force generated during MTs disassembly, Fig. \ref{FigProperties} tells us the {\it critical force} of one MT disassembly is about 1.2 pN 
under the present experimental environment \cite{Akiyoshi2010}. In Fig. \ref{FigPropertiesExtra}(b), the mean growth length $l_g, l_g^*$ and mean shortening length $l_s, l_s^*$ which are given in the definitions of critical force $F_{c10}, F_{c11}$ are also plotted as functions of external force. One can easily see that $l_g\le l_g^*$, and $l_s\le l_s^*$ since the mean dwell time of MT in assembly state $T_g\le1/K_c$ and mean dwell time in disassembly state $T_s\le 1/K_r$. But for large external force, $l_s\approx l_s^*$ since, for such cases, MTs leave disassembly state mainly by rescue.

\begin{figure}
  \includegraphics[width=250pt]{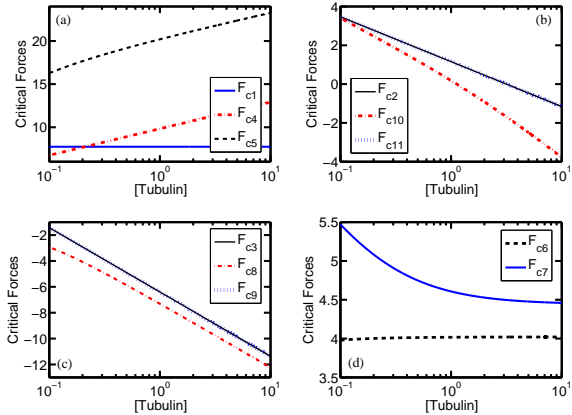}\\
  \caption{{\it Critical forces} as defined in Sec. II.C, in which [Tubulin]=1 means the concentration of GTP-tubulin is the same as the one used in the experiments of Akiyoshi {\it et al} \cite{Akiyoshi2010}. For better understand the curves for $F_{ci}$, see Tab. \ref{table2}.
  }\label{FigCriticalForce}
\end{figure}

\begin{figure}
  \includegraphics[width=250pt]{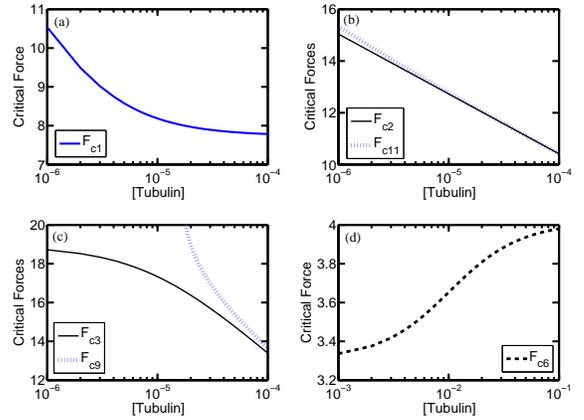}\\
  \caption{Plots of {\it critical forces } $F_{ci}$ for $i=1, 2, 3, 6, 9, 11$  under low GTP-tubulin concentration. The meaning of [Tubulin] is the same as described in the caption of Fig. \ref{FigCriticalForce}.
  }\label{FigCriticalForceDetail}
\end{figure}

\begin{figure}
  \includegraphics[width=250pt]{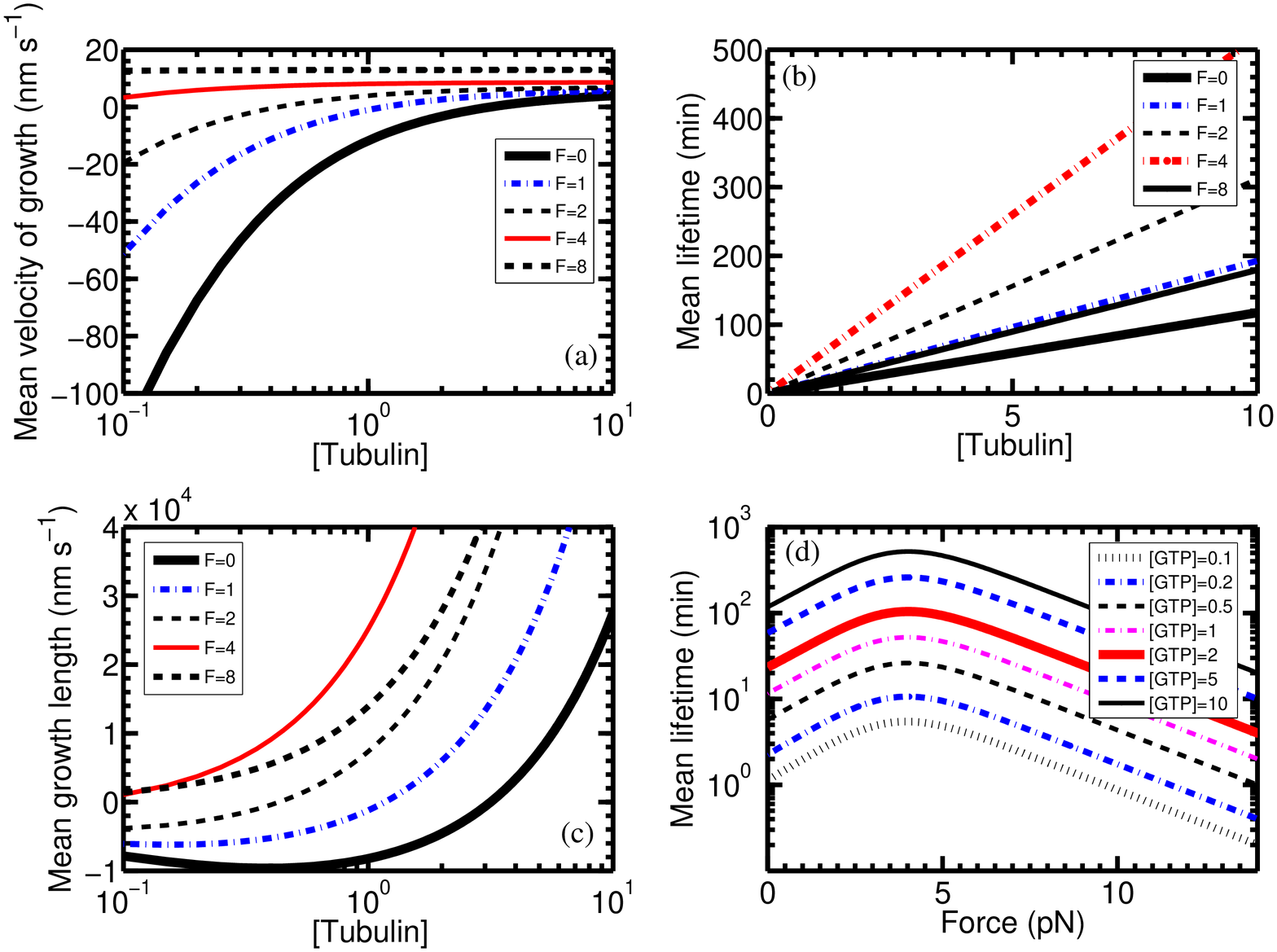}\\
  \caption{GTP-tubulin concentration dependent properties of MT growth and shortening. (a) The mean growth velocity $V$ [see formulation (\ref{eq3})] increases with GTP-tubulin concentration. The plots also indicate $V$ increases with external force [see Fig. \ref{FigProperties}(c)]. (b) and (d) The mean lifetime of bead attachment to MT increases with GTP-tubulin concentration, but increases first and then decreases with external force. For high GTP-tubulin concentration, the critical force $F_{c6}$ under which the mean lifetime gets its maximum is almost a constant (about 4 pN), see also Fig. \ref{FigCriticalForce}(d). (c) The mean growth length of MT before bead detachment does not change monotonically with external force and GTP-tubulin concentration, so there exists critical force under which the maximum is obtained. But for high GTP-tubulin concentration, mean growth length increases with [Tubulin]. (d) For any GTP-tubulin concentration, the mean lifetime does not change monotonically with external force. The optimal value of external force, under which the mean lifetime is maximum, increases with GTP-tubulin concentration, but is almost invariable for large [Tubulin]. (c) (d) Both the mean growth length and mean lifetime do not change monotonically with external force, so there exists optimal values under which the corresponding maximum is reached (see $F_{c6}, F_{c7}$ Fig. \ref{FigCriticalForce}).}\label{FigPropertiesOfATP}
\end{figure}

\begin{figure}
  \includegraphics[width=250pt]{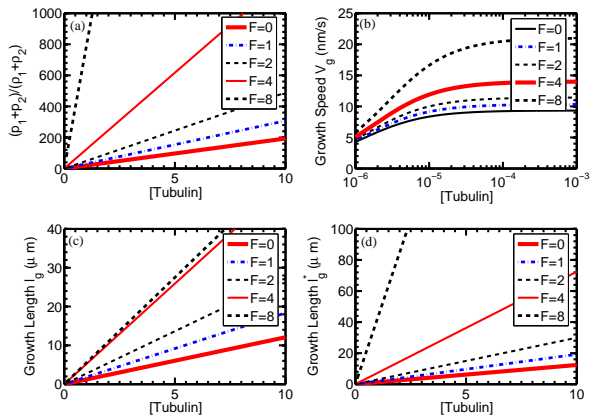}\\
  \caption{The ratio of probability $p_1+p_2$ that MT in assembly state to probability $\rho_1+\rho_2$ that MT in disassembly state, and the mean growth length $l_g, l_g^*$ in each growth period all increase linearly with GTP-tubulin concentration (a) (c) (d) [see Tab. \ref{table3} for definitions]. The growth speed $V_g$ of MT increases with [Tubulin] but tends to a limit constant for saturated concentration [see Eq. (\ref{eq5}) for the formulation of $V_g$]. }\label{FigPropertiesOfATPExtra}
\end{figure}

Since the assembly of MT depends on free GTP-tubulin concentration (in our model, the simple relation $k_1=k_1^0$[Tubulin] is used, and the disassembly process is assumed to be independent of GTP-tubulin concentration, which can be verified by the data in \cite{Walker1988}, see Fig. \ref{FigFittingDataExtra}), the eleven {\it critical forces} defined in the previous section also depend on GTP-tubulin concentration. For convenience, in our calculations (the results are plotted in Figs. \ref{FigCriticalForce} and \ref{FigCriticalForceDetail}), [Tubulin]=1 means the free GTP-tubulin concentration is the same as the one used by Akiyoshi {\it et al} \cite{Akiyoshi2010}.
From Figs. \ref{FigCriticalForce} and \ref{FigCriticalForceDetail}, one can see, the {\it critical forces} $F_{ci}$ for $i=4, 5, 6$ increase, but others decrease with GTP-tubulin concentration [Tubulin]. For high GTP-tubulin concentration, $F_{c2}\approx F_{c11}$ and $F_{c3}\approx F_{c9}$ since for $k_c\ll k_1$, equations $V_g/K_c=V_s/K_r$ and $K_c=K_r$ can be well approximated by $k_1K_r=k_3k_c$ and $k_rk_4(k_1+k_2)
=k_ck_2(k_3+k_4)$ \footnote{If the simple model depicted in Fig. \ref{FigschematicDetachment}(a) is employed to describe the dynamic properties of MT, then $F_{c2}=F_{c11}$ and $F_{c3}=F_{c9}$. The reason is as follows. At steady state, the probabilities $p,\rho$ that MT in assembly and disassembly states are $p=k_r/(k_c+k_r)$ and $\rho=k_c/(k_c+k_r)$ respectively. So the mean growth velocity of MT is $V=k_1p-k_2\rho=(k_1k_r-k_2k_c)/(k_c+k_r)$. Then the {\it critical force} $F_{c2}$ satisfies $k_1(F_{c2})k_r(F_{c2})=k_2(F_{c2})k_c(F_{c2})$. Meanwhile, $l_g^*=V_g/K_c=V_g/k_c=k_1L/k_c$ and $l_s^*=V_s/K_r=V_s/k_r=k_2L/k_r$, so $l_g^*(F_{c11})=l_s^*(F_{c11})$ is equivalent to $k_1(F_{c11})k_r(F_{c11})=k_2(F_{c11})k_c(F_{c11})$ which means $F_{c2}=F_{c11}$. At the same time, $p=\rho$ is equivalent to $k_c=k_r$, so $F_{c3}=F_{c9}$. But for our model as depicted in Fig. \ref{FigschematicDetachment}(b), $F_{c2}\neq F_{c11}$ and $F_{c3}\neq F_{c9}$ [see Figs. \ref{FigCriticalForceDetail}(b) and \ref{FigCriticalForceDetail}(c)].}.
Since the {\it force distribution factors} $\delta_g>0$, $\delta_s<0$ (see Tab. \ref{table1}), from Eqs. (\ref{eq5}) (\ref{eq6}) one can easily show that the growth speed $V_g$ increases but the shortening speed $V_s$ decreases with external force $F$. Therefore, $V_g<V_s$ if $F<F_{c1}$ (see Tab. \ref{table2}). Eqs. (\ref{eq5}) (\ref{eq6}) also indicate that the growth speed $V_g$ increases with but the shortening speed $V_s$ is independent of GTP-tubulin concentration [Tubulin]. Therefore, the {\it critical force} $F_{c1}$ decreases with GTP-tubulin concentration [Tubulin] [see Fig. \ref{FigCriticalForceDetail}(a)]. But for high [Tubulin], {\it critical force} $F_{c1}$ is almost a constant [see Fig. \ref{FigCriticalForce}(a)] since, for saturating concentration, the growth speed $V_g$ tends to a constant [see Eq. (\ref{eq5}) and Fig. \ref{FigPropertiesOfATP}(a)].  The decrease of {\it critical force} $F_{c2}$ with [Tubulin] can be easily seen from expression (\ref{eq17}) [see Fig. \ref{FigCriticalForce}(b)]. The decrease of {\it critical forces} $F_{c1}, F_{c2}$ implies that low GTP-tubulin concentration might be helpful to chromosome segregation. From expressions in (\ref{eq2}) one can verify $(p_1+p_2)/(\rho_1+\rho_2)=k_rk_4[k_1+k_2]/k_ck_2[k_3+k_4]$. So $(p_1+p_2)/(\rho_1+\rho_2)$ increases linearly with [Tubulin] [see Fig. \ref{FigPropertiesOfATPExtra}(a)]. At the same time, $\delta_r+\delta_s>0, \delta_g>0$ and $\delta_g+\delta_c<0, \delta_s<0$ (see Tab. \ref{table1}) imply $(p_1+p_2)/(\rho_1+\rho_2)$ also increases with external force $F$ [see Fig. \ref{FigPropertiesExtra}(a)]. Therefore, the {\it critical force} $F_{c3}$ decreases with [Tubulin] [see Fig. \ref{FigCriticalForce}(c)].

Since the detachment rate $K_a$ increases and detachment rate $K_d$ decreases with external force $F$ [see Fig. \ref{FigFittingData}(a)], and $K_a$ increases with but $K_d$ is independent of [Tubulin] [see Eqs. (\ref{eq13}) (\ref{eq14})], the {\it critical force} $F_{c4}$ increases with [Tubulin] [see Fig. \ref{FigCriticalForce}(a)]. The increase of {\it critical force} $F_{c5}$ indicates MTs will spend more time in assembly state at high GTP-tubulin concentration [see Tab. \ref{table2} and Figs. \ref{FigCriticalForce}(a) and \ref{FigProperties}(b)]. The increase of {\it critical force} $F_{c6}$ [see Fig. \ref{FigCriticalForceDetail}(d)] implies, the peak of the lifetime-force curve as plotted in Fig. \ref{FigFittingData}(d) will move rightwards as the increase of [Tubulin], but with a upper bound around 4 pN [see Figs. \ref{FigCriticalForce}(d) and \ref{FigPropertiesOfATP}(d)]. Similarly, the decrease of {\it critical force} $F_{c7}$ [see Fig. \ref{FigCriticalForce}(d)] means, the peak of the mean growth length-force curve will move leftwards as the increase of [Tubulin], and with lower bound around 4.44 pN. Finally, {\it critical forces} $F_{c8}$, $F_{c9}$, $F_{c10}$, $F_{c11}$ all decrease with [Tubulin]. It may need to say that, in Ref. \cite{Akiyoshi2010}, only experimental data for positive force cases are measured, and similar experimental methods as used in Refs. \cite{Dogterom1997, Janson2003} might be employed to apply negative force to MTs. At the same time, the mechanism of MT growth and shortening under negative external force cases might be completely different from that under positive external force cases, so for the results of {\it critical forces} plotted in Fig. \ref{FigCriticalForce} which have negative values, experimental verification should be firstly done before further analysis.

To better understand the GTP-tubulin concentration [Tubulin] dependent properties of MT assembly and disassembly, more figures are plotted in Figs. \ref{FigPropertiesOfATP} and \ref{FigPropertiesOfATPExtra}. One can see that the mean lifetime $T$, ratio $(p_1+p_2)/(\rho_1+\rho_2)$, and mean growth length $l_g, l_g^*$ all increase linearly with [Tubulin] (from the corresponding formulations, one can easily see that the mean shortening speed $V_s$, and mean shortening length $l_s, l_s^*$ are all independent of [Tubulin]). The mean velocity $V$ and mean growth speed $V_g$ also increase with [Tubulin], but tend to a external force $F$ dependent constant [one can verify this limit constant is $k_2L=k_2^0L\exp(F\delta_g/k_BT)$. For such cases, the MT stays mainly in sub-state 2, i.e., $p_2\approx1$, see Fig. \ref{FigProperties}(a)]. The mean growth length $VT$ does not change monotonically with external force [see Figs. \ref{FigProperties}(a) and \ref{FigPropertiesOfATP}(c)] but increases with [Tubulin] for high GTP-tubulin concentration cases.

\section{Concluding remarks}
In this study, a two-state mechanochemical model of microtubulin (MT) growth and shortening is presented. In assembly (growth) state, one GTP-tubulin will attach to the growing tip of the protofilament (PF) firstly and then, after the hydrolysis of GTP in the penultimate PF unit, the curved PF segment is slightly straightened with one new PF unit lying into the MT cylinder surface. In disassembly (shortening) state, one tubulin unit will detach from the tip of PF, and then the GDP (or GDP+Pi) capped tip segment of PF will be further curved with one new tubulin unit out of the MT surface (the phosphate is assumed to be released simultaneously). The PF can switch between the assembly and disassembly states with external force dependent rates stochastically. Each assembly or disassembly process contributes to one step of growth or shortening of MT with step size $L=$0.615 nm. This model can fit the recent experimental data measured by Akiyoshi {\it et al} \cite{Akiyoshi2010} well.

From this model, interesting properties of MT growth and shortening are found: Under large external force or high GTP-tubulin concentration, the MT is mainly in assembly state; The mean lifetime of bead attachment to MT and mean growth length during this period (in experiments, the external force is applied to MT through a bead attached to the growing tip of MT) increase firstly and then decrease with the external force, but roughly speaking, they all increase with the GTP-tubulin concentration; The growth speed of MT increases with GTP-tubulin concentration but has an external force dependent limit. For the sake of experimental verification, altogether eleven {\it critical forces} are defined, including the force under which the mean lifetime or mean growth length reach its maximum, the mean assembly speed is equal to the mean disassembly speed, the probabilities of MT in assembly and disassembly states are equal to each other, the detachment rates of bead during assembly and disassembly states are the same, the mean dwell times in assembly and disassembly states are the same, the mean growth velocity of MT is vanished, {\it etc}. Almost all of the above {\it critical forces} decrease with the GTP-tubulin concentration, since high GTP-tubulin concentration is favorable for MT growth and under low GTP-tubulin concentration, MT will shortens its length in average. Roughly speaking, GTP-tubulin and external force are helpful to MT assembly, but there exists optimal values external force for the mean lifetime of bead on MT and mean growth length of MT.

\vskip 0.5cm

\acknowledgments{This study is funded by the Natural
Science Foundation of Shanghai (under Grant No. 11ZR1403700). The author thanks Michael E. Fisher of IPST in University of Maryland for his initial introduction and inspiration of the present study, and is also very appreciated for the referees' critical comments and valuable suggestions, due to which many changes have been done.
}

\end{document}